\documentclass[
aps,
twocolumn,
superscriptaddress,
prc,
showkeys,
showpacs,
nofootinbib,
notitlepage,
floatfix,
]{revtex4-2}

\usepackage{xcolor}
\usepackage{amssymb}
\usepackage{graphicx}% Include figure files
\usepackage{subfigure} 
\usepackage{bm}% bold math
\usepackage[T1]{fontenc}
\usepackage{xspace}
\usepackage{booktabs}
\usepackage{tabularx}
\usepackage{array}
\usepackage{siunitx}
\usepackage{dcolumn}
\usepackage{amsmath}
\usepackage[colorlinks=true,citecolor=blue,filecolor=blue,linkcolor=blue,urlcolor=blue,pdftex]{hyperref}
\usepackage{orcidlink} % For ORCID links in author list
\usepackage{currfile}

\usetikzlibrary{shapes,arrows,positioning}

\newcommand{\cevns}{{\text{CE}$\nu$\text{NS}}}
\begin{document}
% \linenumbers  
\title{Low-Energy Nuclear Recoil Calibration of XENONnT with a \texorpdfstring{$^{88}$Y}{Y-88}Be Photoneutron Source}

\newcommand{\bologna}{\affiliation{Department of Physics and Astronomy, University of Bologna and INFN-Bologna, 40126 Bologna, Italy}}
\newcommand{\chicago}{\affiliation{Department of Physics, Enrico Fermi Institute \& Kavli Institute for Cosmological Physics, University of Chicago, Chicago, IL 60637, USA}}
\newcommand{\coimbra}{\affiliation{LIBPhys, Department of Physics, University of Coimbra, 3004-516 Coimbra, Portugal}}
\newcommand{\columbia}{\affiliation{Physics Department, Columbia University, New York, NY 10027, USA}}
\newcommand{\lngs}{\affiliation{INFN-Laboratori Nazionali del Gran Sasso and Gran Sasso Science Institute, 67100 L'Aquila, Italy}}
\newcommand{\mainz}{\affiliation{Institut f\"ur Physik \& Exzellenzcluster PRISMA$^{+}$, Johannes Gutenberg-Universit\"at Mainz, 55099 Mainz, Germany}}
\newcommand{\mpik}{\affiliation{Max-Planck-Institut f\"ur Kernphysik, 69117 Heidelberg, Germany}}
\newcommand{\munster}{\affiliation{Institut f\"ur Kernphysik, University of M\"unster, 48149 M\"unster, Germany}}
\newcommand{\nikhef}{\affiliation{Nikhef and the University of Amsterdam, Science Park, 1098XG Amsterdam, Netherlands}}
\newcommand{\nyuad}{\affiliation{New York University Abu Dhabi - Center for Astro, Particle and Planetary Physics, Abu Dhabi, United Arab Emirates}}
\newcommand{\purdue}{\affiliation{Department of Physics and Astronomy, Purdue University, West Lafayette, IN 47907, USA}}
\newcommand{\rice}{\affiliation{Department of Physics and Astronomy, Rice University, Houston, TX 77005, USA}}
\newcommand{\stockholm}{\affiliation{Oskar Klein Centre, Department of Physics, Stockholm University, AlbaNova, Stockholm SE-10691, Sweden}}
\newcommand{\subatech}{\affiliation{SUBATECH, IMT Atlantique, CNRS/IN2P3, Nantes Universit\'e, Nantes 44307, France}}
\newcommand{\torino}{\affiliation{INAF-Astrophysical Observatory of Torino, Department of Physics, University  of  Torino and  INFN-Torino,  10125  Torino,  Italy}}
\newcommand{\ucsd}{\affiliation{Department of Physics, University of California San Diego, La Jolla, CA 92093, USA}}
\newcommand{\wis}{\affiliation{Department of Particle Physics and Astrophysics, Weizmann Institute of Science, Rehovot 7610001, Israel}}
\newcommand{\zurich}{\affiliation{Physik-Institut, University of Z\"urich, 8057  Z\"urich, Switzerland}}
\newcommand{\paris}{\affiliation{LPNHE, Sorbonne Universit\'{e}, CNRS/IN2P3, 75005 Paris, France}}
\newcommand{\freiburg}{\affiliation{Physikalisches Institut, Universit\"at Freiburg, 79104 Freiburg, Germany}}
\newcommand{\napels}{\affiliation{Department of Physics ``Ettore Pancini'', University of Napoli and INFN-Napoli, 80126 Napoli, Italy}}
\newcommand{\nagoya}{\affiliation{Kobayashi-Maskawa Institute for the Origin of Particles and the Universe, and Institute for Space-Earth Environmental Research, Nagoya University, Furo-cho, Chikusa-ku, Nagoya, Aichi 464-8602, Japan}}
\newcommand{\laquila}{\affiliation{Department of Physics and Chemistry, University of L'Aquila, 67100 L'Aquila, Italy}}
\newcommand{\tokyo}{\affiliation{Kamioka Observatory, Institute for Cosmic Ray Research, and Kavli Institute for the Physics and Mathematics of the Universe (WPI), University of Tokyo, Higashi-Mozumi, Kamioka, Hida, Gifu 506-1205, Japan}}
\newcommand{\kobe}{\affiliation{Department of Physics, Kobe University, Kobe, Hyogo 657-8501, Japan}}
\newcommand{\kit}{\affiliation{Institute for Astroparticle Physics, Karlsruhe Institute of Technology, 76021 Karlsruhe, Germany}}
\newcommand{\tsinghua}{\affiliation{Department of Physics \& Center for High Energy Physics, Tsinghua University, Beijing 100084, P.R. China}}
\newcommand{\ferrara}{\affiliation{INFN-Ferrara and Dip. di Fisica e Scienze della Terra, Universit\`a di Ferrara, 44122 Ferrara, Italy}}
\newcommand{\groningen}{\affiliation{Nikhef and the University of Groningen, Van Swinderen Institute, 9747AG Groningen, Netherlands}}
\newcommand{\westlake}{\affiliation{Department of Physics, School of Science, Westlake University, Hangzhou 310030, P.R. China}}
\newcommand{\shenzhen}{\affiliation{School of Science and Engineering, The Chinese University of Hong Kong (Shenzhen), Shenzhen, Guangdong, 518172, P.R. China}}
\newcommand{\coimbrapoli}{\affiliation{Coimbra Polytechnic - ISEC, 3030-199 Coimbra, Portugal}}
\newcommand{\uniheidelberg}{\affiliation{Physikalisches Institut, Universit\"at Heidelberg, Heidelberg, Germany}}
\newcommand{\roma}{\affiliation{INFN-Roma Tre, 00146 Roma, Italy}}
\newcommand{\bucknell}{\affiliation{Department of Physics \& Astronomy, Bucknell University, Lewisburg, PA, USA}}
\author{E.~Aprile\,\orcidlink{0000-0001-6595-7098}}\affiliation{Physics Department, Columbia University, New York, NY 10027, USA}
\author{J.~Aalbers\,\orcidlink{0000-0003-0030-0030}}\affiliation{Nikhef and the University of Groningen, Van Swinderen Institute, 9747AG Groningen, Netherlands}
\author{K.~Abe\,\orcidlink{0009-0000-9620-788X}}\affiliation{Kamioka Observatory, Institute for Cosmic Ray Research, and Kavli Institute for the Physics and Mathematics of the Universe (WPI), University of Tokyo, Higashi-Mozumi, Kamioka, Hida, Gifu 506-1205, Japan}
\author{S.~Ahmed Maouloud\,\orcidlink{0000-0002-0844-4576}}\affiliation{LPNHE, Sorbonne Universit\'{e}, CNRS/IN2P3, 75005 Paris, France}
\author{L.~Althueser\,\orcidlink{0000-0002-5468-4298}}\munster
\author{B.~Andrieu\,\orcidlink{0009-0002-6485-4163}}\affiliation{LPNHE, Sorbonne Universit\'{e}, CNRS/IN2P3, 75005 Paris, France}
\author{E.~Angelino\,\orcidlink{0000-0002-6695-4355}}\affiliation{INAF-Astrophysical Observatory of Torino, Department of Physics, University  of  Torino and  INFN-Torino,  10125  Torino,  Italy}\affiliation{INFN-Laboratori Nazionali del Gran Sasso and Gran Sasso Science Institute, 67100 L'Aquila, Italy}
\author{D.~Ant\'on~Martin\,\orcidlink{0000-0001-7725-5552}}\affiliation{Department of Physics, Enrico Fermi Institute \& Kavli Institute for Cosmological Physics, University of Chicago, Chicago, IL 60637, USA}
\author{F.~Arneodo\,\orcidlink{0000-0002-1061-0510}}\affiliation{New York University Abu Dhabi - Center for Astro, Particle and Planetary Physics, Abu Dhabi, United Arab Emirates}
\author{L.~Baudis\,\orcidlink{0000-0003-4710-1768}}\affiliation{Physik-Institut, University of Z\"urich, 8057  Z\"urich, Switzerland}
% ...
\collaboration{XENON Collaboration}\email[]{xenon@lngs.infn.it}\noaffiliation
\date{\today}

\begin{abstract}
Characterizing low-energy, keV-range nuclear recoils near the detector threshold is one of the major challenges for large direct dark matter detectors. To that end, we have successfully used an Yttrium-Beryllium photoneutron source that emits 152~keV neutrons for the calibration of the light and charge yields of the XENONnT experiment for the first time. After data selection, we accumulated 474~events from 183~hours of exposure with this source. The expected background was $55 \pm 12$~accidental coincidence events, estimated using a dedicated 152~hour background calibration run with an Yttrium-PVC gamma-only source and data-driven modeling. From these calibrations, we extracted the light (charge) yield for liquid xenon at our field strength of 23\,V/cm between 0.3\,(0.7)\,keV$_{\rm NR}$ and 5.0\,keV$_{\rm NR}$. This calibration is crucial for accurately measuring the solar $^8$B neutrino coherent elastic neutrino-nucleus scattering and searching for light dark matter particles with masses below 12~GeV/c$^2$. 
\end{abstract}


\maketitle

\section{\label{sec:intro}Introduction}
\begin{figure}[t]
    \centering
	\includegraphics[width=\columnwidth]{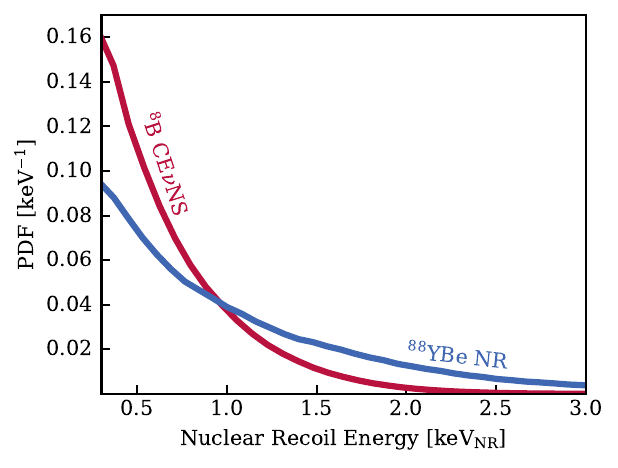}
	\caption{Comparison of the single-scatter nuclear recoil energy probability density function (PDF) between coherent elastic neutrino-nucleus scattering from solar $^{8}$B neutrinos (red) and $^{88}$YBe neutrons entering the TPC (blue). The similarity and large overlap makes $^{88}$YBe neutrons an ideal source to calibrate signal yields for the $^8$B neutrino coherent elastic neutrino-nucleus scattering. 
    }
	\label{fig:ybe_b8_recoil_energy_com}
\end{figure}

The past decade witnessed major improvements in xenon-based dark matter~(DM) experiments, bringing us to the verge of entering the ``neutrino fog''~\cite{OHare:2021utq}. $^{8}$B neutrinos from the Sun transfer O(1\,keV) energy to the entire nucleus of xenon via a process called coherent elastic neutrino-nucleus scattering (\cevns)~\cite{Freedman:1973yd,Kopeliovich:1974mv,XENON:2024ijk}. Recently, solar $^{8}$B \cevns~signals were first indicated by the liquid xenon time projection chambers (TPCs) of XENONnT~\cite{XENON:2024ijk} and PandaX~\cite{PandaX:2024muv} close to the three-sigma level. Such a low-energy signal is also characteristic of DM with a mass around 6 GeV/$c^2$, favored by some light DM models~\cite{Kaplan:2009ag,Tulin:2017ara, light_wimp}. To avoid mismodelling these low-energy nuclear recoil~(NR) signatures, it is therefore imperative to properly calibrate the detector response to them. To that end, we conducted an NR calibration of the XENONnT TPC~\cite{XENON:2024wpa} using an external Yttrium-Beryllium ($^{88}$YBe) photoneutron source. As shown in Fig.\,\ref{fig:ybe_b8_recoil_energy_com}, the neutrons from the $^{88}$YBe source that enter the TPC produce a recoil energy spectrum similar to $^{8}$B \cevns, providing the opportunity to characterize this rare process. 

XENONnT is one of the leading DM detectors searching for weakly interacting massive particles (WIMPs). It uses a dual-phase TPC filled with liquid xenon, of which 5.9 tonnes constitute its active target~\cite{XENON:2024wpa}. 

Particle interactions inside the TPC generally produce both scintillation light and ionization electrons in the detector. While the scintillation light generates a prompt signal (S1), the ionization electrons drift upwards and are extracted into a gas phase at the top of the TPC, via applied electric fields. Drift time is the duration required for electrons to drift to the liquid-gas interface, with the maximum drift time being 2.2 ms for XENONnT. The extracted electrons produce a secondary scintillation signal (S2) due to their interaction with xenon atoms in the gas phase. Both the S1 and S2 signals are detected in two arrays of photomultiplier tubes (PMTs) placed on the top and bottom of the detector. The drift time between the S1 and S2 signals is used to determine the depth of the interaction which is defined as Z-coordinate, and the scintillation light pattern of the S2 signal as recorded with the top PMT array is used to reconstruct the X and Y coordinates of the interaction. Position and drift time are then used to correct for detector effects and obtain the \textit{corrected} cS1 and cS2 signal strengths~\cite{XENON:2024qgt}. Furthermore, the ratio between the areas of the S1/S2 signals is crucial for discriminating between NR events and electronic recoil (ER) events. Particles such as neutrons and WIMPs cause NR events by interacting with xenon nuclei, while particles like gamma-rays and beta particles cause ER events by interacting with electrons in xenon atoms~\cite{XENON:2024xgd}.

This paper describes the calibration of XENONnT using a $^{88}$YBe neutron source~\cite{Collar:2013xva}, achieving a low-energy calibration down to ($0.30 \pm 0.02$)~keV$_{\rm NR}$. In the following, we will report the kinetic energy of the neutron in keV and the energy of the nuclear recoil in keV$_{\rm NR}$. To understand the background arising from $^{88}$Y gamma emissions in the source, additional data was taken with the neutron-emitting beryllium replaced by polyvinyl chloride (PVC) plastic of comparable density and shape. 

\begin{figure*}[htbp]
    \centering
    \raisebox{10mm}{% Adjust as needed
        \includegraphics[scale=1.21]{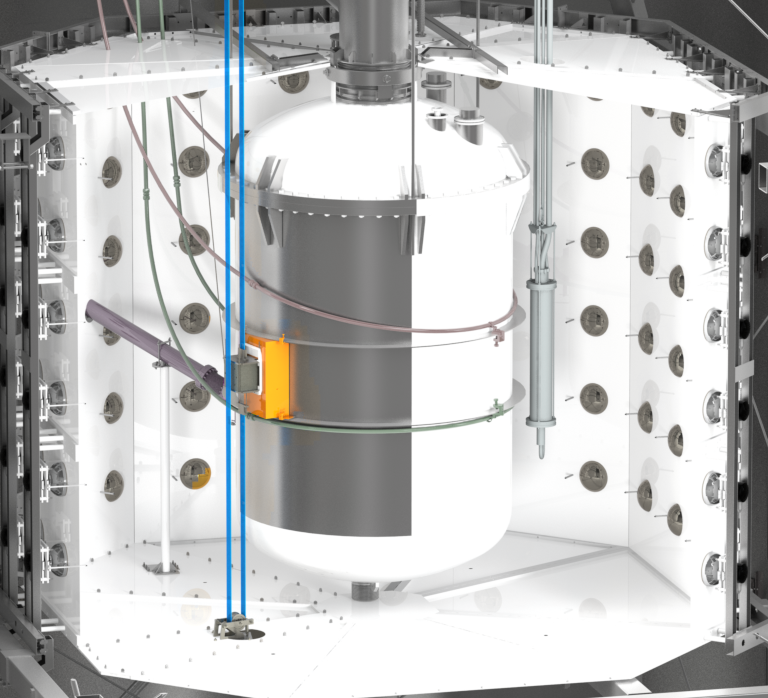}
    }
    \hspace{1mm}
    \includegraphics[scale=0.88]{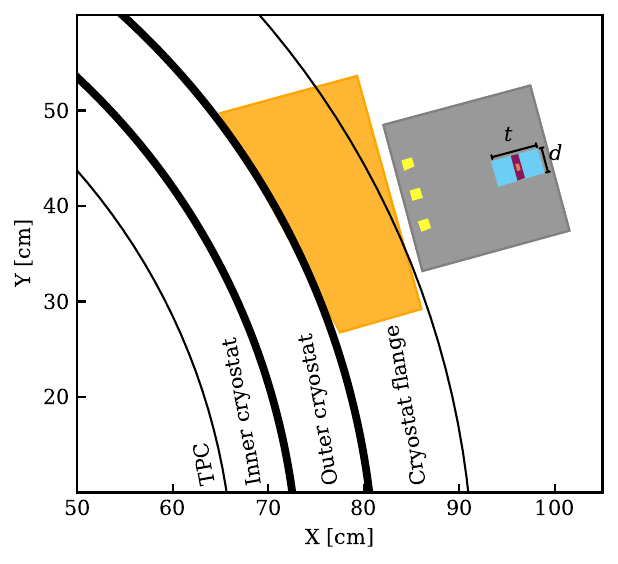}
    \caption{Left: A CAD rendering of the XENONnT cryostat inside the neutron veto detector (see~\cite{XENON:2024wpa} for more details). Also shown is the I-belt system (blue), along with the tungsten source box (grey) and the water displacement box (orange). Right: The $^{88}$YBe source assembly, along with the relevant cryostat parts. In addition to the orange stainless steel air-box, used to displace water between the TPC and the source, and grey tungsten source box, yellow indicates air holes that exist due to the manufacturing process of the tungsten shield, light-blue represents beryllium, purple indicates the acrylic disc that the $^{88}$Y source is made of, and the small orange point indicates the $^{88}$Y activity. The geometric parameters being optimized are the diameter (d) and thickness (t) of the cavity in the tungsten block containing the beryllium target and $^{88}$Y source~\cite{Qin2023}.}
    \label{fig:ibelt_source_geometry}
\end{figure*}

In the following sections, we will first introduce the design and preparation of the $^{88}$YBe calibration in Sec.~\ref{sec:ybe_source}. Following that, we will discuss calibration operations and explain how to evaluate various backgrounds when selecting data in Sec.~\ref{sec:event_selction}, and finally proceed with fitting and extraction the light and charge yields in Sec.~\ref{sec:fitting}.

\section{\label{sec:ybe_source}Design of The \texorpdfstring{$^{88}$Y}{Y-88}Be source}

\subsection{Design Considerations and Challenges}
The $^{88}$YBe photoneutron source produces quasi-monoenergetic 152\,keV neutrons via photodisintegration of $^9$Be by 1.84 MeV $\gamma$-rays from $^{88}$Y decays~\cite{Collar:2013xva}. 

One of the key challenges of calibrating with such a source is the low neutron rate due to the small $^9$Be($\gamma,n$)$^8$Be cross-section of $0.65\,\mathrm{mb}$~\cite{Robinson:2016kxn}, requiring overwhelming $\gamma$-ray activity. This low cross section corresponds to a mean free path between photodisintegration interactions of approximately $\rho_{\text{Be}}/m_{\text{Be}}/(0.65\,\mathrm{mb})\approx 125 \, \text{m}$, highlighting the comparatively high gamma activity needed to construct a useful calibration source.

As meter-scale liquid xenon TPCs have relatively long electron drift times at the millisecond-scale~\cite{Aprile2006-vk, Hunter1988LowenergyED}, a similarly long event window for matching S1 and S2 signals is required. As a result, the rate of mismatched S1 and S2 signals is high for such detectors. These background events, called accidental coincidence (AC) events, can reach up to 1\,kHz without dedicated treatment due to the high activity of the $^{88}$Y source.
In addition, ER events can contaminate the NR calibration region of interest due to their high rate, even after discrimination between ER and NR events is considered~\cite{XENON:2023cxc}. 
Therefore, a high atomic number material is needed to effectively block most of the $\gamma$-rays from the $^{88}$Y decays, while minimizing the energy loss of neutrons passing through it.

 We mounted the source assembly on the I-belt system of XENONnT~\cite{XENON:2024wpa}, and lowered it into the neutron veto detector hosted inside the water Cherenkhov muon veto detector~\cite{XENON:2024wpa}. The I-belt subsystem is shown in Fig.~\ref{fig:ibelt_source_geometry} (left) in blue, and allows us to control the vertical position of the source and lower it into position beside the cryostat during calibration runs.

The size of the assembly that houses the $^{88}$YBe source is restricted to a $(16~\rm cm)^3$ cube due to mechanical clearances in the I-belt system. This motivates the use of tungsten as the shielding material instead of the more commonly-used lead~\cite{Collar:2013xva, LZ:2024bsz}; while tungsten has a lower atomic number, the density of bulk tungsten is higher, leading to a lower radiation length~\cite{ParticleDataGroup:2022pth}.

Lowering the source assembly past the vacuum flange at the top of the cryostat requires it to be set back with respect to the outer diameter of the cryostat, moving it $\sim$10\,cm away from the cryostat's outer wall. To avoid the broadening of the neutron energy spectrum and the reduction of the neutron flux, we designed and installed an air-filled stainless steel box to displace the water between the TPC and the calibration source with air. The air box was manufactured and fixed to the exterior of the cryostat of XENONnT before commissioning. 

\subsection{ Optimization of the Source Geometry}
In addition to tackling the design challenges mentioned earlier, and to ensure a successful calibration, we optimized the geometry of the source assembly using simulations to improve the signal-to-background ratio. A schematic of the full $^{88}$YBe source assembly is shown in Fig.~\ref{fig:ibelt_source_geometry}. The $^{88}$YBe source is composed of a $^{88}$Y disc source, containing a $\rm 5~mm\times3~mm~(diameter\times height)$ active element, sandwiched between two identical, cylindrical Be targets. 

We used Geant4~\cite{GEANT4:2002zbu} simulations to understand the trade-off between increasing the beryllium target size to increase neutron yields and the corresponding decrease in the amount of tungsten shielding needed to accommodate a larger beryllium target while not exceeding the maximum total physical size of $(16~\rm cm)^3$, thereby allowing us to improve the signal-to-background ratio. 
These simulations were used to determine the optimal diameter and thickness of the beryllium targets, as indicated in Fig.~\ref{fig:ibelt_source_geometry}, using a figure of merit based on the ratio of the rate of nuclear recoil to that of electronic recoil events. 

To streamline the calibration procedure, a stainless steel capsule with $2\sim3$ mm thickness was added to the design to contain the $^{88}$YBe source and the beryllium targets. Due to the increased amount of shielding material that needs to be removed to accommodate this stainless steel capsule, the chosen dimensions were slightly smaller than the optimal dimensions derived from simulations; however, the reduced size of the beryllium target was not found to result in a difference that was statistically significant given our simulation statistics. The final source assembly has dimensions $50~{\rm mm}\times25.4~{\rm mm}$ ($t \times d$) excluding the steel capsule, and $55.5~{\rm mm}\times29.4~{\rm mm}$ ($t \times d$) including the steel capsule, resulting in a final $94.5\,\mathrm{mm}$ of tungsten between the source capsule and the TPC. More details regarding the optimization procedure can be found in Appendix~\ref{appendix:optimization}.

\begin{figure*}[thbp]
    \centering
	\includegraphics[width=\textwidth]{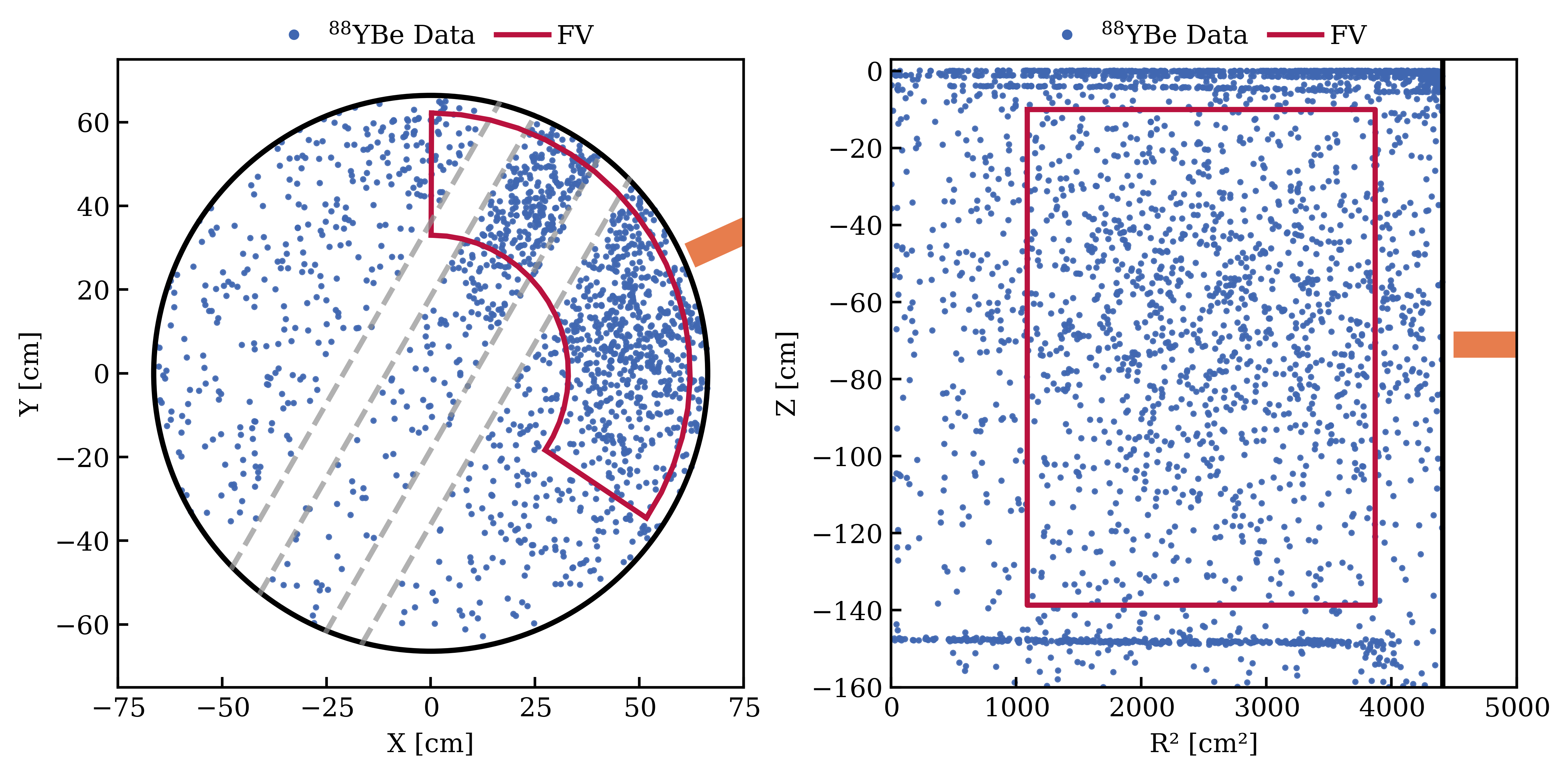}
	\caption{$^{88}$YBe event (blue dots) distribution shown in the XY (left) and R$^{2}$Z (right) planes. The X, Y and Z coordinates represent the reconstructed position of the largest S2 signal in the event. The events projected onto the XY plane have been selected based on a cut along the z-axis, specifically within the range of -139 cm $\lesssim$ z $\lesssim$ -10 cm. The black solid lines show the position of TPC walls and the orange solid lines indicate the position of the source. The region within the red lines is the employed fiducial volume (FV). Four grey dashed lines indicate two regions near perpendicular wires, which were excluded from the fiducial volume. Clustered events at the top of the R$^{2}$Z plane are surface background events, while those at the bottom are cathode events. Events which are outside the fiducial volume and uniformly distributed in the XY plane are mainly residual AC events after removing events in periods with a high rate of delayed electrons and isolated S1 signals.} 
	\label{fig:ybe_fv}
\end{figure*}

\section{\texorpdfstring{$^{88}$Y}{Y-88}Be Calibration and Event Selection}\label{sec:event_selction}

\subsection{Calibration Operations}

$^{88}$YBe calibration operations started with the commissioning of the I-belt system, followed by data-taking with the beryllium targets, and finally data-taking with a PVC dummy target replacing the beryllium target. All components that entered the water tank were cleaned with ethanol and deionised water to minimize any impact on the transparency of the water. During commissioning, a waterproof camera was lowered into the water to verify the operation of the I-belt system and that all mechanical components had the expected mechanical clearance. 

After these tests, the tungsten source box was brought back to the water surface, the source assembly containing the targets and the $^{88}$Y disc source were inserted, and the combined assembly was lowered back into the calibration position shown in Fig.~\ref{fig:ibelt_source_geometry} (left). Starting from September 30$\rm ^{th}$ 2022, $^{88}$YBe source data was taken till October 9$\rm ^{th}$, accounting for a total livetime of approximately 183 hours. The $^{88}$Y
$\gamma$-source had an activity of 316\,kBq at the start of the calibration. Following this $^{88}$YBe calibration, the $^{9}$Be target was replaced by a PVC target with the exact same geometry, and data was taken for an effective livetime of around 152 hours from 10$\rm ^{th}$ October 2022. During this calibration, the detector was operated at a drift field of $(23 \pm 1.5)$~V/cm, the same as in the science search runs~\cite{XENON:2024ijk,XENON:2023cxc}.

\subsection{Event Selection}
Following the data acquisition, comprehensive analyses were performed to select NR events and subsequently determine the light and charge yields. Due to moderation by the materials in the path, predominantly the residual water between the source and the air box in the neutron veto, the mean kinetic energy of neutrons entering the TPC was lower than the emission energy of 152\,keV, with the energies mainly ranging from 10\,keV to 100\,keV. Since neutrons tend to scatter multiple times in the large volume of liquid xenon and 80\% of simulated $^{88}$YBe NR events had more than one reconstructed S2, MS neutron NR events were included in the event selection. This was done to maximize statistics and to probe the lowest possible NR energy, even though the geometry of the source had been optimized on the ratio of SS NR events to $\gamma$-induced ER background. Selected events had to have at least 2 observed photons and no more than 10 observed photons in the S1 signal. This upper bound was motivated by the scarcity of NR events with more than 10 observed photons in the S1 signal, as only 7\% of NR events were expected to exceed this threshold. Since the mean free time between two scatters for 100 keV neutrons is approximately 30 ns, prompt scintillation photons from different scatters within a MS NR event are merged into one S1. Consequently, in MS NR events, scatters which generate 0 or 1 observed scintillation photon can also be included in this selection. In this way, the energy threshold of this calibration can be further lowered compared to what is obtained from only selecting SS NR events, which were required to have at least 2 observed photons in the S1 signal. For clarity, the event selection described below was based on the largest S2 in the event, while the S1 corresponded to the total prompt scintillation photons from all scatters combined. 

Selecting NR events from the $^{88}$YBe source data is challenging, because both $\gamma$-rays and neutrons emitted from the $^{88}$YBe source interact within the liquid xenon. While neutrons deposit energy through elastic scattering with xenon nuclei, $\gamma$-rays primarily interact with xenon atomic electrons via Compton scattering. In this analysis, $^{88}$YBe neutron NR events do not overlap with $\gamma$-ray ER events, as NR events deposit much lower detectable energies (few keV) and the cS2/cS1 leakage from $\gamma$-ray ER events is negligible due to the source design. However, $\gamma$-ray ER events with large energy depositions can also induce delayed electrons~\cite{XENON:2021qze} and isolated S1 signals. Delayed electrons can appear after signal detection, even beyond the maximum drift time. Possible origins for delayed electrons include electrons captured by impurities in the liquid xenon and released later~\cite{XENON:2021qze}, or electrons trapped at the liquid-gas interface~\cite{Akimov:2016rbs}. Isolated S1 signals refer to S1 signals which are not paired with S2 signals within one maximum drift time and may originate from pile-up of PMT afterpulses or PMT dark counts. Isolated S1 signals and delayed electrons can lead to the formation of AC events in two ways, becoming a major background in the low-energy region of NR events. Firstly, the AC events can arise from a random pairing of S2 signals from delayed electrons and isolated S1 signals. Secondly, isolated S1 signals can also form a match with true NR S2 signals, thereby producing AC events. A high rate of delayed-electron S2 signals was observed in the region of $\rm S2 < 120$ PE ($\lesssim$ 4 electrons), due to the increased probability of coinciding delayed electrons within the same S2 time window, leading to a significant population of AC events. This motivated the lower S2 area bound of 120 PE, while the upper bound of 900 PE was set to minimize ER leakage from $\gamma$-ray ER events.

\begin{figure}[htbp]
    \centering
	\includegraphics[width=\columnwidth]{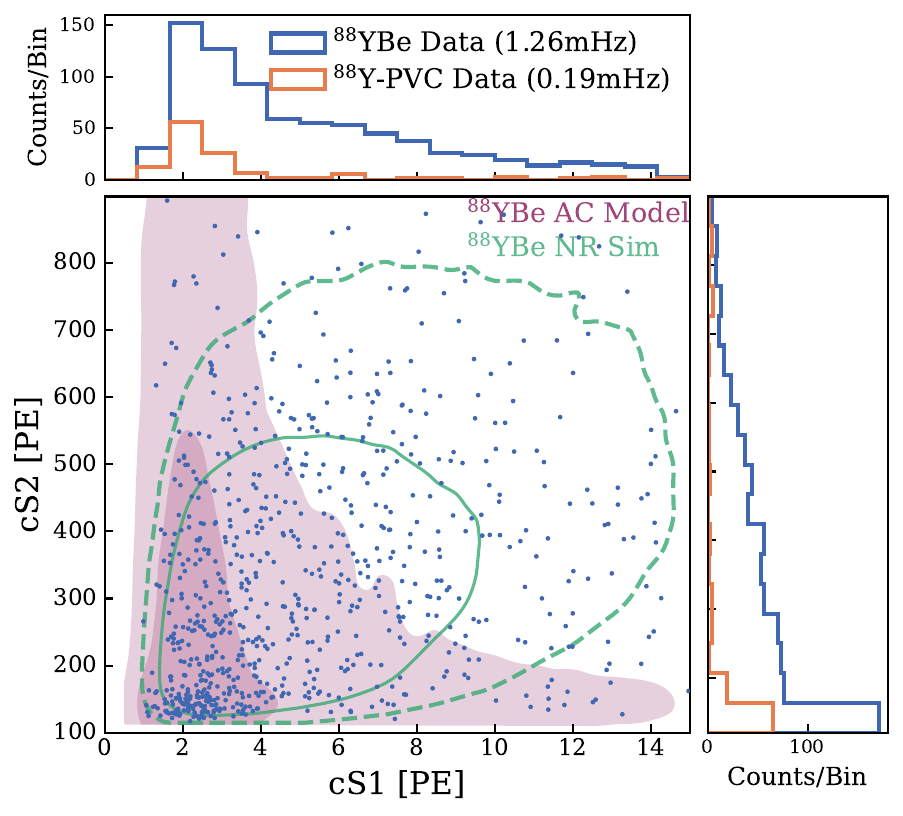}
	\caption{$^{88}$YBe events are represented by blue scatter points in the space of the corrected S1 signal and the corrected largest S2 signal, denoted as cS1 and cS2, respectively. The dark (light) purple region represents the 68\% (95\%) expected AC events as predicted by the $^{88}$YBe AC model. Additionally, the solid (dashed) green contour delineates the 68\% (95\%) expected NR events. A clustering of events within cS2 < 200 PE and cS1 < 4 PE shows a substantial population of AC events. 1D projections in the cS1 and cS2 space are shown for both $^{88}$YBe (blue) and $^{88}$Y-PVC (orange) calibration data, displayed adjacently. The histogram of the $^{88}$Y-PVC data is scaled by a factor of 1.4 to account for the differences in calibration time and the activity of the $^{88}$Y source. The $^{88}$YBe data includes both AC and NR events, whereas the $^{88}$Y-PVC data contains only AC events. Projections for the $^{88}$Y-PVC data reveal a significant presence of AC events, particularly in the lowest energy bins, which also suggests a substantial quantity of AC events in the $^{88}$YBe data. }
	\label{fig:ybe_y88_compare}
\end{figure}

\begin{figure}[htbp]
    \centering
	\includegraphics[width=\columnwidth]{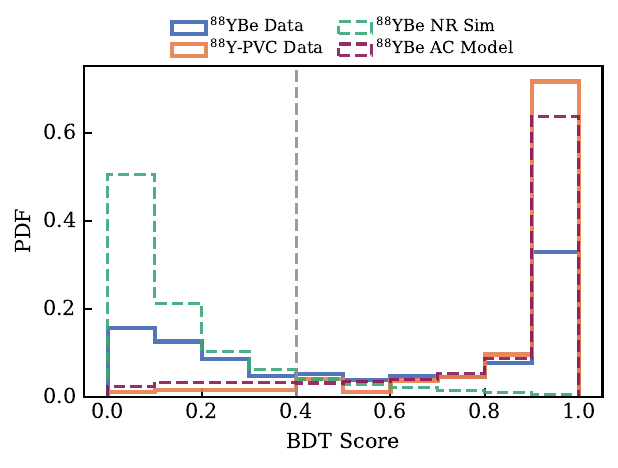}
	\caption{BDT score distribution, the probability of an event being an AC event, as predicted by the trained BDT classifier. A smaller BDT score indicates a higher probability of an event being NR, while a larger BDT score suggests a higher probability for an event being AC. In $^{88}$Y-PVC data, all events are assumed to be AC events while $^{88}$YBe data contains both NR and AC events. The BDT score distributions of NR and AC simulation are very distinct, which reflects the satisfactory discrimination power of the trained model. In addition, the grey dashed line represent the cut threshold.}
	\label{fig:ybe_ms_bdt}
\end{figure}

It was observed that delayed electrons and isolated S1 signals exhibit significantly higher rates following large signals, with delayed electrons also showing positional dependence on large preceding S2 signals. To reduce the AC background, we first removed events that were temporally and spatially close to S1 signals larger than 1000 PE or S2 signals larger than 10000 PE, as detailed in~\cite{XENON:2024ijk}.

\begin{figure*}[htbp]
    \centering
    \subfigure{%
        \includegraphics[width=\columnwidth]{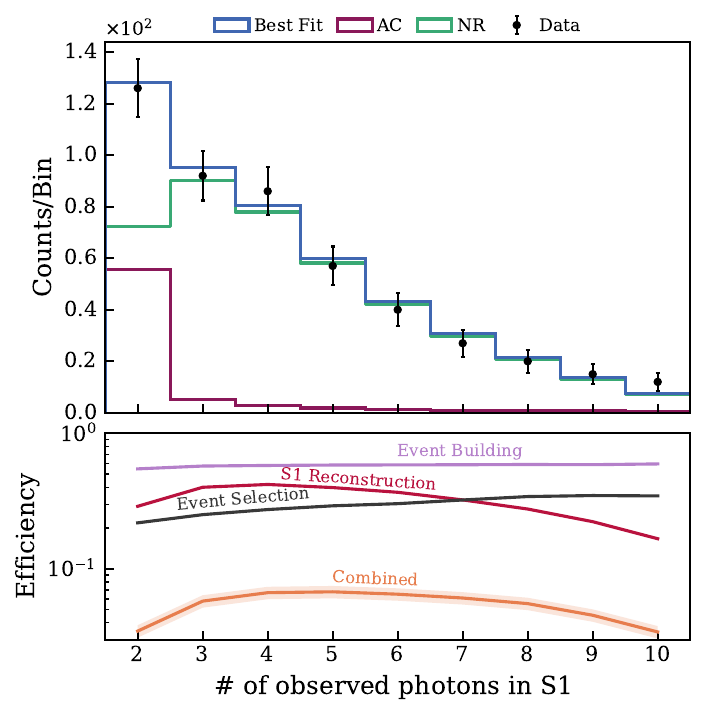}
    }
    \hfill
    \subfigure{%
        \includegraphics[width=\columnwidth]{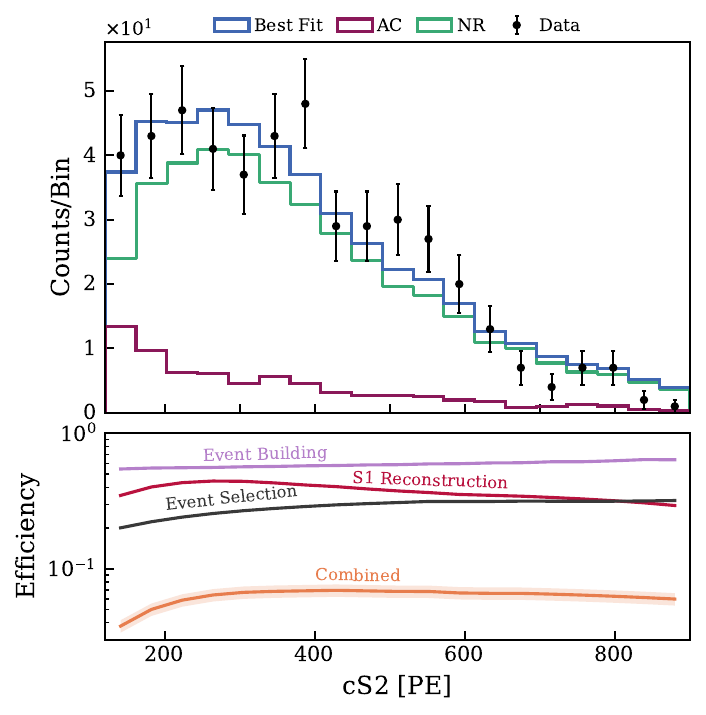}
    }
    \caption{Top: Comparison of $^{88}$YBe best fit simulation (blue) to the $^{88}$YBe data (black) in terms of the number of observed photons in S1 (left) and the largest cS2 (right). Both comparisons exhibit excellent agreement between the simulation and the data. Additionally, the individual contributions of Nuclear Recoil (NR) events (green) and Accidental Coincidence (AC) events (purple) are depicted. Bottom: 1D projections of the combined efficiency (orange), which is the product of the S1 reconstruction (red), event building (violet), and event selection efficiency (dark grey), are shown with respect to the number of observed photons in S1 (left) and cS2 (right).}
    \label{fig:ybe_sim_data_compare}
\end{figure*}

Two additional cuts were developed to further improve the signal-to-background ratio. The first was a optimized fiducial volume cut. As the mean free path for neutrons with these low energies entering the TPC is approximately $10\,\mathrm{cm}$~\cite{Bell:2021ihi,Shibata:2011JENDL} in the liquid xenon, most NR events occur near the TPC wall and are spatially correlated with the source position. However, in the $^{88}$YBe data, there was an additional spatially uniform component, similar to that in $^{88}$Y-PVC data, which originates from AC events. This component can be reduced using an optimized fiducial volume. The shape of the fiducial volume was determined by a $^{88}$YBe NR-only simulation, aiming to encompass 90\% of simulated NR events. Additonally, the drift field is distorted~\cite{XENONnT:2023dvq} near the TPC wall. This leads to the effect that ionization electrons produced from interaction sites close to the wall can be captured by the TPC wall during the drifting to the liquid-gas interface, in turn reducing the size of observed S2 signals originating from regions near the walls. ER events from the $^{88}$Y $\gamma$-rays were used to study the loss of S2 signal strength near the TPC wall. This analysis resulted in the adoption of an outer radius of 62.2 cm for the fiducial volume, thereby limiting the events adversely affected by this effect to 7\%. While this charge loss effect is modeled in the analysis, this choice of fiducial volume helps avoid issues arising from mis-modeling at radii where this effect becomes dominant. In addition, two regions near perpendicular wires on the gate and anode electrodes, which were employed to prevent electrode sagging, were also excluded due to the known difficulty~\cite{XENONnT:2023dvq} of modeling the pulse shape of S2 signals in these regions. The final shape of the fiducial volume is shown in Fig.~\ref{fig:ybe_fv}, where it can be seen that the fiducial volume cut removed a large fraction of AC events. 

Events after all above-mentioned cuts are presented in the cS1 and cS2 space in Fig.~\ref{fig:ybe_y88_compare}, which are the S1 and largest S2 signals corrected for their positions inside the TPC~\cite{XENON:2024qgt}.  Comparing the $^{88}$Y-PVC and $^{88}$YBe data shows that the majority of selected events are NR events. However, the residual events in the $^{88}$Y-PVC data indicate a significant presence of AC events in the region of interest (ROI), as the majority of $\gamma$-ray ER events have S2 signals larger than 900 PE, while the remaining events in the ROI arise from the random pairing of isolated S1 signals and delayed electrons induced by $\gamma$-rays. 

To better understand AC events, we construct data-driven AC model \textit{Axidence}~\cite{axidence}, following a systematic procedure. First, we prepare S1 and S2 ensembles. The S1 ensemble includes all isolated S1 signals, defined as S1 signals that are not paired with any S2 signal. The S2 ensemble includes both isolated S2 signals, defined as S2 signals that are either not paired with any S1 signal, and S2 signals associated with S1 $\leq10$ observed photons. Here, isolation refers only to the absence of a partner of the opposite type: additional S1 signals may still accompany an isolated S1 signal, and additional S2 signals may accompany S2 in the prepared S2 ensemble. These prepared S1 and S2 signals, together with their surrounding S1 and S2 signals, are resampled using a bootstrap technique and processed through the full reconstruction pipeline. During the resampling stage, an AC event is formed either when an isolated S1 and an S2 from the ensembles fall within the maximum drift time of 2.2\,ms, or when an isolated S1 replaces the original small S1 associated with an S2. In both cases, the resulting event is identified as an AC event by the reconstruction algorithm. All event-selection criteria applied in the main analysis are also applied to these simulated AC events to ensure consistency. Finally, the AC model is normalized using the measured rates of isolated S1 signals and S2 signals in the S2 ensemble (i.e., isolated S2s and S2s accompanied by S1 $\leq$10 photons) after all cuts, together with the total livetime of the calibration dataset, in order to reproduce the expected AC background rate. The robustness of the AC model methodology is validated using the $^{88}$Y–PVC dataset. After applying all analysis cuts except the final BDT cut, 88 events remain in the $^{88}$Y–PVC data, compared to 72 events predicted by the AC model. This level of agreement, within approximately 20\%, demonstrates that the model reliably reproduces the rate of AC events. We therefore assign a conservative 20\% systematic uncertainty to the AC prediction in the $^{88}$YBe dataset.

We rely on the distinction between MS NR events and AC events to improve our signal-to-noise. Depending on the separation in space and time among scatters, one or multiple S2 signals can be reconstructed from MS NR events. For events with more than one reconstructed S2 within the event time window, the time, position, and pulse shape information of the two largest S2 signals were used to distinguish between MS and AC events, as the time and position differences among scatters in MS events are correlated. NR events with only one reconstructed S2 were difficult to distinguish from AC events and were excluded from the data selection. A boosted decision tree (BDT) classifier~\cite{Chen:2016:XST:2939672.2939785} was employed to enhance the discrimination between NR events and AC events, thereby further reducing the AC background. To train the BDT classifier, the $^{88}$YBe NR-only simulation~\cite{WFSim} was used as signal. The \textsuperscript{88}Y-PVC AC model served as the background for training the BDT model. 

To evaluate the performance of the trained BDT model, an independent testing set with one-fourth of the size of the training set was used. After training, the model demonstrated a true positive (signal) rate of 92\% and a true negative (background) rate of 86\% on the test set, indicating a robust performance with no signs of overfitting. 
The final selection cut was defined based on the BDT score after training, which represents the predicted probability of an event being AC. The BDT score distribution is shown in Fig.~\ref{fig:ybe_ms_bdt}. The cut was set to exclude events with an AC score greater than 0.4, effectively rejecting 91\% of AC events while retaining 89\% of NR events. After applying all cuts, a total of 474 events remained in the $^{88}$YBe data. However, a fraction of AC events was still expected to exist among the selected events, particularly those arising from the random pairing of S2 signals from NRs and isolated S1 signals, which do not exist in the $^{88}$Y-PVC data. Therefore, another data-driven AC model~\cite{XENON:2024xgd, XENON:2024ijk} based on the $^{88}$YBe data was developed to estimate the number of the remaining AC events after the BDT cut, yielding an expectation of $(55 \pm 12)$ AC events within the dataset.

\section{Modeling and Fit Results\label{sec:fitting}}

In order to determine the low-energy NR light and charge yields, quantifying the number of photons and electrons generated per keV$_{\rm NR}$ of energy deposition in liquid xenon, the selected $^{88}$YBe events were fitted by the detector response model using Appletree~\cite{appletree}. The fit region was defined to be the number of observed photons in S1 $\in$ [2, 10] and cS2 $\in$ [120, 900] PE. The detector response model, as detailed in Ref.~\cite{XENON:2024xgd}, comprises both the yield model and the detector reconstruction model to generate NR signals. The parameterized Noble Element Simulation Technique (NEST v2)~\cite{nestv2} yield model is employed to converts deposited energy into scintillation photons and ionization electrons. Subsequently, the detector reconstruction model transforms these produced photons and electrons into the observed S1 and S2 signals. We will detail our procedures as follows: yield model, event efficiencies, NR signal modeling, and parameter extraction\,(fitting). 

In NEST v2, the parameters \(\alpha\) and \(\beta\) control the total number of quanta produced by an NR. The drift field dependence of yields in NEST v2 is controlled by two parameters, $\gamma$ and $\delta$, with the relationship $\varsigma \propto \gamma \mathcal{E}^{\delta}$, where $\varsigma$ denotes the field dependence and $\mathcal{E}$ represents the drift field strength. However, the $^{88}$YBe calibration data was acquired at a single drift field strength of $(23 \pm 1.5)$ V/cm. In the fit, $\delta$ was fixed at its NEST v2 default value of -0.0533 to avoid degeneracy. The parameters \(\epsilon\), \(\zeta\), and \(\eta\) control the shape of the charge yield, while \(\theta\) and \(\iota\) control the shape of the light yield. The parameters \(F_{ex}\) and \(F_{i}\) determine the statistical fluctuations of excitons and electron-ion pairs, respectively, while \(\xi\) and \(\omega\) govern the variance of recombination. We kept all parameters' priors in the yield model, except for $\delta$, uniformly distributed within specific ranges for each parameter. The prior distributions for each parameter in the yield model are detailed in Table~\ref{tab:para_fit_nestv2}. 

\begin{figure*}[htbp]
    \centering
	\includegraphics[width=0.99\textwidth]{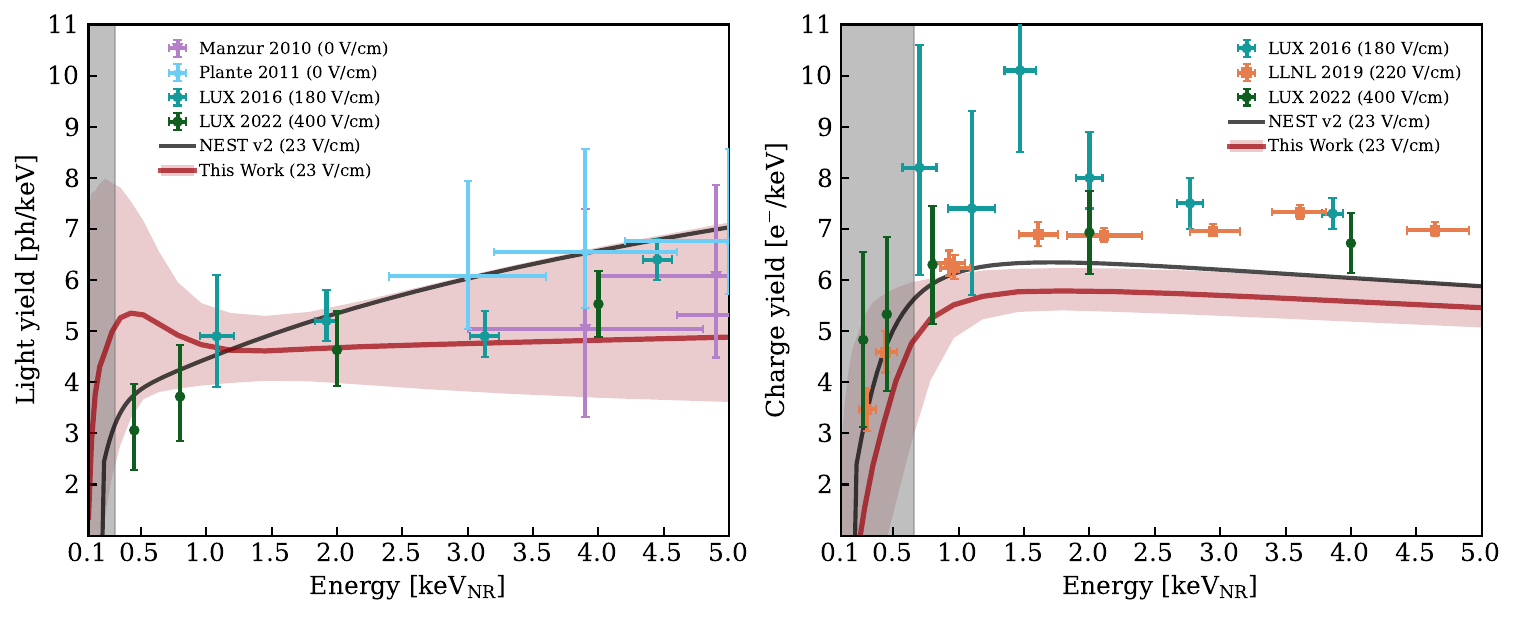}
	\caption{Light yield~(left) and charge yield~(right) extracted from the fit to the $^{88}$YBe calibration data. The red solid curves represent the nominal yields and red bands show $\pm$1$\sigma$ uncertainties of yields. The uncertainty bands are calculated using sampled values of fit parameters in the last 10,000 iterations. The yields from NEST v2.3.11 are shown in black solid curves for comparison. Measurements from other experiments~\cite{Lenardo:2019fcn, LUX:2016ezw, LUX:2022qxb, Manzur:2009hp, Plante:2011hw}, taken at different drift field strength, are also shown for comparison. The yields in the grey shaded region are presented solely for completeness, as the lowest energies to which this calibration is sensitive are $(0.30 \pm 0.02)~\text{keV}_{\rm NR}$ for the light yield and $(0.66 \pm 0.04)~\text{keV}_{\rm NR}$ for the charge yield.}
	\label{fig:ybe_fitted_yields}
\end{figure*}

\renewcommand{\arraystretch}{1.3}

\begin{table}[hbtp]
\caption{Prior and marginal posterior distribution of each parameter in the NR yield model from the $^{88}$YBe calibration. The default values of each parameter from NEST v2.3.11 are also shown as a comparison.}
\centering
\addtolength{\tabcolsep}{-0.1cm}
\begin{tabular}{c c c c
        }

\hline
\textbf{Parameter} & \textbf{NEST v2} & \textbf{Prior}& \textbf{Marginal posterior} \\
\hline
$\alpha$ & $ 11.0 \substack{+2.0 \\ -0.5}$ & [8.5, 21] & $ 12.1 \substack{+2.7 \\ -2.3}$ \\
$\beta$ & $ 1.10 \substack{+0.05 \\ -0.05}$ & [0.85, 1.35] & $ 1.01 \substack{+0.14 \\ -0.13}$ \\
$\gamma$ & $ 0.0480 \substack{+0.0021 \\ -0.0021}$ & [0.0375, 0.0585] & $ 0.0466 \substack{+0.0072 \\ -0.0065}$ \\
$\epsilon$ & $ 12.6 \substack{+3.4 \\ -2.9}$ & [0, 29.6] & $ 16.3 \substack{+6.8 \\ -6.1}$ \\
$\zeta$ & $ 0.3 \substack{+0.1 \\ -0.1}$ & [0, 0.8] & $ 0.4 \substack{+0.3 \\ -0.3}$ \\
$\eta$ & $ 2 \substack{+1 \\ -1}$ & [-3, 7] & $ 4 \substack{+2 \\ -2}$ \\
$\theta$ & $ 0.30 \substack{+0.05 \\ -0.05}$ & [0.05, 0.55] & $ 0.31 \substack{+0.17 \\ -0.18}$ \\
$\iota$ & $ 2.0 \substack{+0.5 \\ -0.5}$ & [-0.5, 4.5] & $ 1.8 \substack{+1.9 \\ -1.8}$ \\
$F_{ex}$ & 0.4 & [0, 1] & $ 0.5 \substack{+0.3 \\ -0.3}$ \\
$F_{i}$ & 0.4 & [0, 1] & $ 0.5 \substack{+0.3 \\ -0.3}$ \\
$\xi$ & 0.50 & [0, 1] & $ 0.51 \substack{+0.34 \\ -0.34}$ \\
$\omega$ & 0.19 & [0, 1] & $ 0.48 \substack{+0.34 \\ -0.33}$ \\
\hline
\end{tabular}

\label{tab:para_fit_nestv2}
\end{table}

Besides the yield model, modeling the detector reconstruction of  low-energy NR events is crucial for obtaining reliable yields. Various efficiencies were studied. Firstly, MS NR events have a lower S1 reconstruction efficiency compared to SS NR events. Due to the low speed ($\sim0.01c$) of $^{88}$YBe neutrons and their ability to scatter multiple times in liquid xenon, the time separation between single photons in S1 signals can be larger than that from a SS event. This can lead to incorrect classification of S1 signals as S2 signals or unknown (neither S1 nor S2), thereby reducing the S1 reconstruction efficiency for MS NR events. This effect was more pronounced for MS NR events with large S1 signals, which result from the merging of photons produced by a large number of scatters. Therefore, a map for the S1 reconstruction efficiency was built from a large sample of simulated MS NR events and subsequently employed in the fitting process. Secondly, the high rate of delayed electrons and isolated S1 signals caused by $\gamma$-rays can reduce the efficiency of event building, as the event building algorithm may fail when physical S1 or S2 signals are surrounded by too many signals comparable in size. To estimate the event building efficiency for $^{88}$YBe NR events, a salting study using Saltax~\cite{saltax} was performed by injecting simulated NR events into $^{88}$YBe calibration data. If simulated NR events are introduced into periods with a high rate of delayed electrons and isolated S1 signals, the event building efficiency for these simulated NR events will be reduced. This method enabled assessing how delayed electrons and isolated S1 signals influence event building, yielding a measured efficiency of 57.4\%. Thirdly, delayed electrons exhibit strong spatial correlations with $\gamma$-ray-induced events, which predominantly occur near the TPC wall. The spatial correlations introduce a spatial dependence in the event selection efficiency, which was evaluated by the study of injecting simulated NR events into the data as well. For instance, the event selection efficiency at R = 60 cm was only 25\% of that at R = 40 cm, thereby illustrating this spatial dependence. The combined efficiency was the product of S1 reconstruction efficiency, event building efficiency and event selection efficiency, as illustrated in Fig.~\ref{fig:ybe_sim_data_compare}. The uncertainty in the combined efficiency was conservatively estimated to be 20.6\%, primarily attributed to the uncertainties associated with the S1 reconstruction efficiency and event selection efficiency.

The \textsuperscript{88}YBe NR signals were simulated using the detector response model implemented in Appletree~\cite{appletree} as follows. The energy deposition and positional information for each scatter were obtained from Geant4 simulations. These Geant4 simulations modeled both the transport of neutrons through the detector materials, and the position and energy of nuclear recoils produced from the interactions between neutrons and the liquid xenon target. The energy depositions of these nuclear recoils are then clustered at the microphysics scale using the epix package~\cite{XENON:2024xgd, daniel_wenz_2023_7516942}.

The yield model was then applied to simulate the number of scintillation photons, \(N_\text{ph}\), and ionization electrons, \(N_\text{e}\). The number of observed scintillation photons from each energy deposition were determined by \(N_\text{ph}\), the photon gain (\(g_{1} = 0.137 \pm 0.001~\text{PE/photon}\))\, the double photon emission probability (\(p_\text{DPE} = 0.229 \pm 0.024\)), and the single photon acceptance (\(p_\text{sp} = 0.921 \pm 0.010\)), with the total number of observed photons in the S1 signal for a MS NR event being the sum over all energy depositions~\cite{XENON:2024xgd}. Electrons from distinct energy depositions were merged when they occurred in close temporal and spatial proximity, as indicated by full waveform simulations, leading to one or more reconstructed S2 signals for a MS NR event~\cite{XENON:2024qgt, XENON:2024xgd, WFSim}. The area of each S2 signal was determined by the electron lifetime (\(\tau = 21.8 \pm 0.2~\text{ms}\)) and the electron gain (\(g_{2} = 16.8 \pm 0.4~\text{PE/electron}\)). The reconstructed position of each S2 signal was computed as a weighted average of the positions of the contributing energy depositions, where the weights corresponded to the number of drifted electrons from each contributing energy deposition. The largest S2 signal used in the fitting was then corrected for position and electron lifetime effects to yield the cS2 signal~\cite{XENON:2024qgt}. With the simulated number of observed photons in the S1 signals and the cS2 signals for MS NR events, the previously described efficiencies were applied. We assigned a 5\% S1 efficiency, conservatively estimated by the salting method. The S2 efficiency is estimated based on the acceptance of the same set of cuts. The efficiencies are estimated based on calibration and sideband data, with values taken to be conservative.
A normalization factor, \(N_{\mathrm{nr}} =  (1.6 \pm 1.4) \times 10^4
\), was applied to scale the NR signal simulation to the expected number of YBe-induced NR events. It was derived by normalizing the simulated NR rate to the observed \(^{129}\mathrm{Xe}(n,n^\prime\gamma)\) 39.6\,keV activation line in data. The dominant uncertainty on this normalization arises from the modeling of the xenon neutron-capture \((n,\gamma)\) cross section, because subsequent capture gammas can overshadow the 39.6\,keV activation line and thereby change its visible rate. This systematic was evaluated by comparing the baseline Geant4 configuration, \texttt{QGSP\_BERT\_HP}, with alternative physics-list and nuclear-data treatments for neutron transport and interactions~\cite{Geant4Toolkit,Geant4PhysicsListGuide}. The resulting variation in the balance between \(^{129}\mathrm{Xe}(n,n^\prime)^{129m}\mathrm{Xe}\) production and subsequent neutron capture was found to dominate the spread in the inferred \(N_{\mathrm{nr}}\).
The data-driven AC model established previously was used as the background with an additional fit normalization factor (\(N_\text{ac} = 55 \pm 12\)) introduced to scale the AC model without altering the shape of its event spectrum.

In total there were 21 parameters in the detector response model, including 12 yield model parameters and 9 Gaussian-constrained detector reconstruction parameters. An affine invariant Markov chain Monte Carlo algorithm~\cite{foreman2013emcee,Foreman-Mackey2019} was used to sample from this hyperdimensional space to find the best-fit parameters which maximize the posterior. In Fig.~\ref{fig:ybe_sim_data_compare}, the best-fit model shows excellent agreement with the data, yielding a p-value of 0.995 using a binned Poisson $\chi^{2}$ test~\cite{Baker:1983tu, gof} defined by Eq.~\eqref{eq:chisquare}. $\chi^{2}$ is defined to be the log likelihood ratio between the simulation and the data, where $\mathcal{L}$ denotes the likelihood function given by Eq.~\eqref{eq:likelihood}. Here, $y$ and $n$ denote the simulation and the experimental data, respectively. Specifically, $y_{i}$ and $n_{i}$ represent the number of events in the $i^{th}$ bin of the simulation and data in the fit region, respectively.
\begin{align}
\chi^{2} = - 2\ln \mathcal{L}(y;n) + 2\ln \mathcal{L}(n;n) \label{eq:chisquare}\\
\mathcal{L}(y;n) = \prod_i \exp(-y_{i})\cdot y_{i}^{n_{i}}/n_{i}! \label{eq:likelihood}
\end{align}
Additionally, the best-fit values for all detector reconstruction parameters lie within the $\pm$1$\sigma$ uncertainties of the prior estimates.

In Fig~\ref{fig:ybe_fitted_yields}, the constrained yields from this analysis are compared to the default NEST v2 model. The light yield and charge yields from the fit are in agreement with NEST v2 model within $\pm$1$\sigma$ uncertainties. The uncertainties on the fitted light and charge yields were estimated by sampling the fit parameters from the final 10,000 iterations of the fitting procedure. These samples effectively approximate the joint posterior distribution of the parameters. For each set of sampled parameter values, we ran the simulation to obtain the corresponding light and charge yields. The resulting distributions of these yields were then used to determine the $\pm1\sigma$ uncertainties. The larger uncertainty on the light yield arises from two independent effects. First, the light collection efficiency is $\sim$4 times lower than the electron collection efficiency, so obtaining the same number of detected photons requires many more primaries than for electrons. This results in enhanced statistical fluctuations, leading to broader uncertainties in the inferred light yield compared to the charge yield. Second, in the NEST v2 framework the light yield is parameterized with two additional shape parameters, $\theta$ and $\iota$, compared to the charge yield. Given our limited statistics, these extra degrees of freedom are only weakly constrained, and this additional flexibility in the light yield model further broadens its uncertainty. In addition, we find that the low-energy shape of $L_y$ is most sensitive to the parameters $\theta$ and $\zeta$, with variations in other yield parameters producing comparatively smaller
effects.

\begin{figure}[htbp]
    \centering
	\includegraphics[width=\columnwidth]{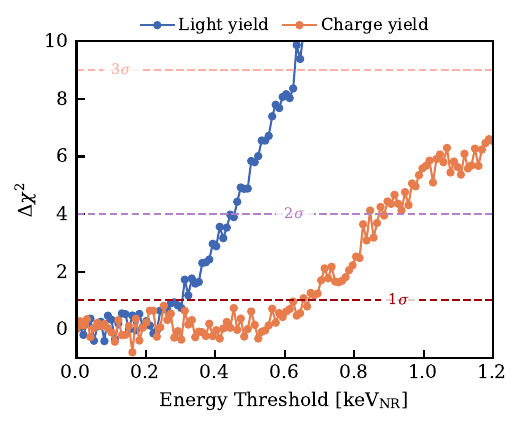}
	\caption{The estimation of the lowest observable energy in the \textsuperscript{88}YBe calibration is illustrated. Each blue (orange) point represents the \(\Delta \chi^2\) value obtained by comparing a template generated with the light (charge) yield set to zero below the corresponding energy threshold on the \(x\)-axis to the best-fit template without any energy threshold. The energy threshold is defined where \(\Delta \chi^2 = 1\), resulting in $(0.30 \pm 0.02)~\text{keV}_{\rm NR}$ for the light yield and $(0.66 \pm 0.04)~\text{keV}_{\rm NR}$ for the charge yield.}
	\label{fig:ybe_energy_threshold}
\end{figure}

To determine the lowest energy sensitivity of the calibration, we used the method described in~\cite{LUX:2022qxb}. A range of energy thresholds $E_{thr}$ are applied in the best-fit detector response model, and the resulting $\chi^{2}(E_\mathrm{thr})$ is compared to $\chi^{2}(0)$ of the model without energy threshold to get $\Delta \chi^{2} = \chi^{2}(E_\mathrm{thr}) - \chi^{2}(0)$ as shown in Fig.~\ref{fig:ybe_energy_threshold}. The energy thresholds of this analysis were found where $\Delta \chi^{2} = 1$. It was found that the lowest observable recoil energies in the \textsuperscript{88}YBe calibration are $(0.30 \pm 0.02)~\mathrm{keV_{nr}}$ for the light yield and $(0.66 \pm 0.04)~\mathrm{keV_{nr}}$ for the charge yield. The threshold $E_{\rm thr}$ is the energy at which the data no longer remain
statistically consistent with a model in which the light (or charge) yield is forced to zero below that value. Thus, $E_{\rm thr}$ is not sensitive to how tightly the yields are constrained in the fit.
The lower threshold observed for the light-yield measurement arises from the multiplicity of scatters in MS NR events. On average, an NR event contains about six scatters; although each individual scatter may produce only $\leq 1$ observed photon, their contributions add coherently within the S1 window. For illustration, a $0.3~\mathrm{keV_{nr}}$ energy deposition corresponds
to a predicted light yield of $L_{y} \simeq 5.1~\mathrm{ph/keV_{nr}}$, or approximately two produced photons. With a 14\% light-collection efficiency, the probability of observing at least one photon from such a scatter is $\sim 26\%$. When several scatters occur in the same event, the
probability that the combined S1 contains at least two detected photons increases substantially, allowing the S1 signal to satisfy the event-selection criterion even when most individual scatters fall below this level. This multi-scatter accumulation reduces the effective energy threshold for the light-yield measurement.
In contrast, the largest S2 signal in an NR event typically originates from only one or two nearby scatters, which
must collectively produce at least four electrons to pass selection. Because the charge signal is dominated by fewer scatters, it does not benefit from the same multiplicity enhancement as S1 signal, leading to a slightly higher threshold for the charge-yield measurement despite its higher
collection efficiency.

\section{Conclusion}

The XENONnT collaboration successfully conducted the first calibration of the XENONnT TPC using low-energy NR events near the detector threshold from a $^{88}$YBe photoneutron source. The application of machine learning methods combined with refined simulations enabled us to maximize the number of reconstructed low-energy NR events and address the analysis challenges posed by the high rate of $\gamma$-ray emissions from the source. This calibration enabled us to constrain the light (charge) yield in liquid xenon down to 0.30 (0.66)~keV$_{\rm NR}$ under a low drift-field condition of 23 V/cm. The constrained yields show no discrepancies with the NEST v2 model and provide vital information for searches involving low-momentum transfer interactions, such as $^{8}$B \cevns~\cite{XENON:2024ijk, PandaX:2024muv}, light WIMPs~\cite{light_wimp}, and other rare low-energy NR interactions. 

Photoneutron sources, such as $^{88}$YBe, offer the advantage of producing quasi-monoenergetic neutrons with relatively low emitted kinetic energies, without the technical complexity of operating a deuterium-deuterium neutron gun. 

As future underground DM detectors also function as solar neutrino observatories, precise knowledge of the yields near the detector threshold will become even more important for scientific discoveries. 

\begin{acknowledgments}
We gratefully acknowledge support from the National Science Foundation, Swiss National Science Foundation, German Ministry for Education and Research, Max Planck Gesellschaft, Deutsche Forschungsgemeinschaft, Helmholtz Association, Dutch Research Council (NWO), Fundacao para a Ciencia e Tecnologia, Weizmann Institute of Science, Binational Science Foundation, Région des Pays de la Loire, Knut and Alice Wallenberg Foundation, Kavli Foundation, JSPS Kakenhi, JST FOREST Program, and ERAN in Japan, Tsinghua University Initiative Scientific Research Program, DIM-ACAV+ Région Ile-de-France, and Istituto Nazionale di Fisica Nucleare. This project has received funding/support from the European Union’s Horizon 2020 research and innovation program under the Marie Skłodowska-Curie grant agreement No 860881-HIDDeN.

We gratefully acknowledge support for providing computing and data-processing resources of the Open Science Pool and the European Grid Initiative, at the following computing centers: the CNRS/IN2P3 (Lyon - France), the Dutch national e-infrastructure with the support of SURF Cooperative, the Nikhef Data-Processing Facility (Amsterdam - Netherlands), the INFN-CNAF (Bologna - Italy), the San Diego Supercomputer Center (San Diego - USA) and the Enrico Fermi Institute (Chicago - USA). We acknowledge the support of the Research Computing Center (RCC) at The University of Chicago for providing computing resources for data analysis.

We thank the INFN Laboratori Nazionali del Gran Sasso for hosting and supporting the XENON project.
\end{acknowledgments}
\bibliography{ybe_paper}% Produces the bibliography via BibTeX.

\clearpage
\appendix

\section{Detailed Source Optimization Procedure}\label{appendix:optimization}

Simulating the $^{88}$YBe source response in Geant4 is computationally expensive because of the low photodisintegration cross-section. We thus used two separate simulations to estimate the event rate in the XENONnT TPC due to $\gamma$-rays and neutrons. First, we estimated the rate of $\gamma$-ray-induced neutrons by simulating 1.84~MeV $\gamma$-rays from the active element. We then estimated the observed neutron rates given the rate of photodisintegration reactions by treating the $^9$Be as an isotropic $152\,\mathrm{keV}$ neutron source~\cite{PhysRev.76.611},

taking light and charge yields, and detector efficiencies~\cite{XENON:2023cxc,XENON:2024wpa, szydagis_2022_6028483} into account. Furthermore, S1 and S2 signals generated from multiple scatters of the same primary particle, but close in time and position, may be merged into single S1 and S2 signals. Such effects were also accounted for in the simulation framework.

As these Geant4 simulations considered the neutron and gamma simulations independently without modeling for the photodisintegration reaction that produces neutrons, we need to compute this probability of neutron emission separately to understand the effective exposure time of the neutron simulation. We did this using a Monte Carlo simulation, where we simulated $10^5$ $\gamma$-rays uniformly from the small cylindrical active element. We then computed the probability of a $^9$Be($\gamma,n$)$^8$Be reaction for each $\gamma$-ray based on the trajectory length of the $\gamma$-ray in the beryllium target ($\ell$) and the number density of beryllium atoms ($N_{\rm Be}$), as $p_i = \sigma_{\gamma, n} \ell N_{\rm Be}$. The average probability of $^9$Be($\gamma,n$)$^8$Be reactions per $\gamma$-ray is then given by
\begin{equation}\label{eq:gamma-n-prob}
    p_{\gamma, n} = \frac{1}{10^5}\sum_i p_i.
\end{equation}

While only multi-scatter (MS) NR events are considered in the final analysis (see Sections~\ref{sec:event_selction} and~\ref{sec:fitting}), during the design stage, we used the ratio of the single-scatter (SS) NR events to the total TPC event rate as the figure of merit (FOM) for the optimization of the source geometry in order to maximize the signal-to-background ratio of the calibration. These SS events are defined as events with one S1 and one S2 signal in the TPC after combining neighboring signals in space-time, as opposed to multi-scatter events with multiple S1 or S2 signals attributed to the same neutron. The classification of events into SS and MS is dependent on the ability to resolve individual peaks.

While in the Geant4 simulation the beryllium elements are considered to be uniform sources, in reality the gamma flux is higher close to the active element of the $^{88}$Y source. To account for this in the computation of the FOM, a per-interaction weight was used to compute the rate of SS NR events in the TPC. This is given by
\begin{equation}\label{eq:simulation_weights}
    w_i = \frac{\mathcal{N}}{r_i^2},
\end{equation}
where $\mathcal{N}$ is a normalizing constant to ensure that the weights sum up to the total number of events, and $r_i$ is the distance between the point of origin of the neutron in the beryllium target and the $^{88}$Y source. Since the dimension of the $^{88}$Y active element is much smaller than the beryllium target sizes being considered, the $^{88}$Y source was approximated as a point source when computing weights. The number of ER or SS NR events is thus computed as
\begin{equation}\label{eq:rate_computation}
    N = \sum_{i \in F} w_i,
\end{equation}
where $F$ is the set of events being considered, such as SS NR events or ER events.

We finally computed the optimization metric using Eqs.~\eqref{eq:gamma-n-prob} and \eqref{eq:simulation_weights} as
\begin{equation}
    \mathcal{S} = \frac{p_{\gamma, n} N_{\rm NRSS}/S_{n}}{p_{\gamma, n} N_\mathrm{ER}/S_{n} + N_{\gamma}/S_{\gamma}},
\end{equation}
where $N_\mathrm{NRSS}$ is the number of SS NR events from the Geant4 neutron simulation after weighting computed using~\eqref{eq:rate_computation}, $N_\mathrm{ER}$ is the number of ERs from the Geant4 neutron simulation after weighing computed using~\eqref{eq:rate_computation}, $N_{\gamma}$ is the number of ERs from the gamma simulation, and $S_{n}$ and $S_{\gamma}$ is the total number of neutron and gamma primaries from the respective simulations, regardless of whether they formed detectable events in the TPC. This optimization metric can be interpreted as the number of SS NR events expected for each background ER event; a value of $\mathcal{S}=1$, for example, would mean that we would expect an equal number of SS NR events and background ER events.

\begin{figure}[thbp]
    \centering
	\includegraphics[width=\columnwidth]{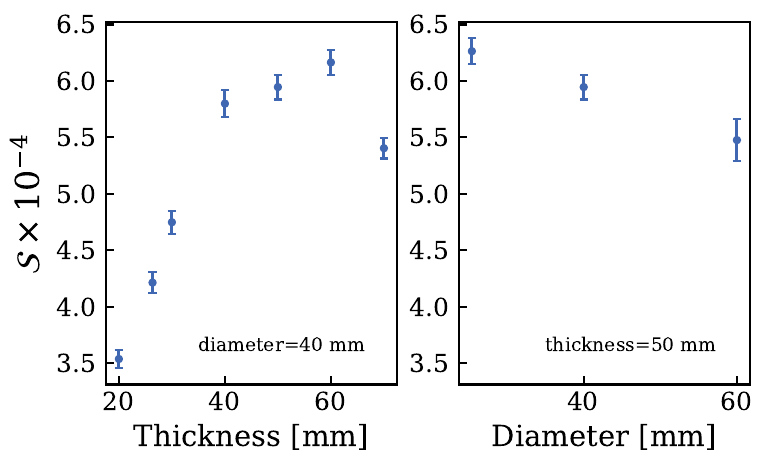}
	\caption{The variation of $\mathcal{S}$ for thickness and diameter optimization of the beryllium target around the $^{88}$Y source. The fixed diameter and thickness values for respective optimizations are also indicated.} 
	\label{fig:ybe_fom_geant4}
\end{figure}

Using $\mathcal{S}$ as the FOM, we optimized the dimensions of the cylindrical beryllium targets in two steps: 1) optimizing the thickness while fixing the diameter at $d=40$ mm, and 2) optimizing the diameter while fixing the thickness at $t=50$ mm. While a full 2D optimization would allow us to be more confident of finding a global minimum, this two-step procedure was chosen due to the computational expense of running a large number of Geant4 simulations. It can be seen in Fig.~\ref{fig:ybe_fom_geant4} that among the considered configurations, the SS NR metric was highest at a thickness of $t=60$ mm and $d=25.4$ mm, the diameter of the $^{88}$Y disc source. To avoid removing too much shielding material when including the thickness of the stainless steel capsule, the chosen dimensions were $t=50$ mm and $d=25.4$ mm. This decision had minimal impact on the resulting neutron yield.

Simulations with the chosen geometry gave us a ratio between the SS NR event rate and the total event rate of $\mathcal{S} = (6.3\pm0.1)\times10^{-4}$, which was higher than all other considered designs shown in Fig.~\ref{fig:ybe_fom_geant4}. In addition to the above geometry, a stainless steel capsule of 2 mm thickness on all sides, and with a screw-on cap of 3 mm thickness was machined to contain the $\rm 50~mm\times25.4~mm$ ($t \times d$) neutron source. The full source assembly, including the stainless steel capsule, had dimensions of $55.5~{\rm mm}\times29.4~{\rm mm}$ ($t \times d$). This resulted in a final $94.5\,\mathrm{mm}$ of tungsten between the source capsule and the TPC.

\end{document}